\documentclass[twocolumn,showpacs,preprintnumbers,prb,amsmath,amssymb,superscriptaddress,floatfix]{revtex4} 

\usepackage{graphicx}
\usepackage{dcolumn}
\usepackage{bm}

\begin{document} 

\setlength{\topmargin}{0in}

\title{New Ground State of  Relaxor Ferroelectric Pb(Zn$_{1/3}$Nb$_{2/3}$)O$_3$
} 
\author{Guangyong Xu}
\affiliation{Physics Department, Brookhaven National Laboratory, Upton, 
New York 11973}
\author{Z.~Zhong} 
\affiliation{National Synchrotron Light Source, Brookhaven National 
Laboratory, Upton, New York 11973}
\author{Y.~Bing}
\affiliation{Department of Chemistry, Simon Fraser University, Burnaby, 
British Columbia, Canada, V5A 1S6}
\author{Z.-G.~Ye}
\affiliation{Department of Chemistry, Simon Fraser University, Burnaby, 
British Columbia, Canada, V5A 1S6}
\author{C.~Stock}
\affiliation{Department of Physics, University of Toronto, Toronto, Ontario, 
Canada M5S 1A7} 
\author{G.~Shirane} 
\affiliation{Physics Department, Brookhaven National Laboratory, Upton, 
New York 11973}
\date{\today} 
 
\begin{abstract} 
High energy x-ray diffraction measurements on 
Pb(Zn$_{1/3}$Nb$_{2/3}$)O$_3$ (PZN)
single crystals show that the system does not have a rhombohedral
symmetry at room temperature as previously believed. The new phase (X) in the 
bulk of the crystal gives Bragg peaks similar to that of a 
nearly cubic lattice with a slight tetragonal distortion.
The Bragg profile remains sharp with no evidence
of size broadening due to the polar micro crystals (MC). 
However, in our preliminary studies of the skin, we have found the expected 
rhombohedral (R) phase as a surface state. 
On the other hand, studies on an electric-field 
poled PZN single crystal clearly indicate a rhombohedral phase at room 
temperature. 

\end{abstract} 
 

\maketitle

\section{Introduction}

\begin{figure}
\includegraphics[width=\linewidth]{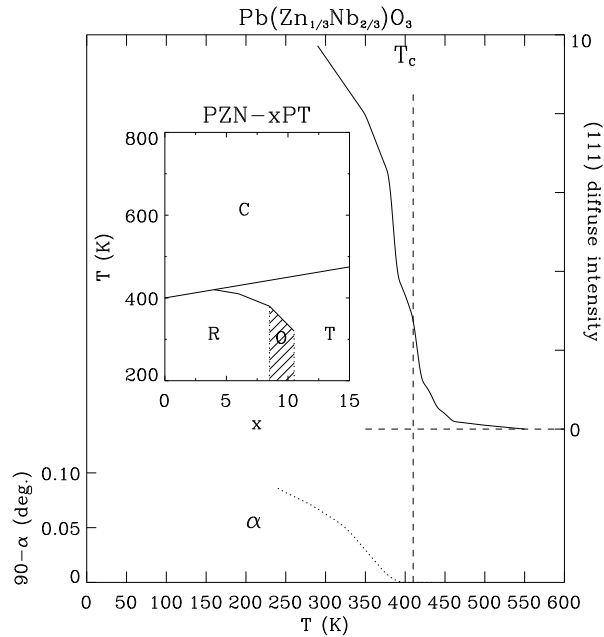}
\caption{Structural properties of PZN. The dotted line in 
the lower left shows the a schematic of the 
rhombohedral angle $\alpha$ vs temperature obtained by Lebon {\it et al.}~\cite{Lebon};
the solid line in the upper right corner shows intensities of neutron 
diffuse scattering  (Ref.~\onlinecite{PZN_diffuse}). 
The inset shows the recently published  
phase diagram in the vicinity of the MPB (Ref.~\onlinecite{PZN_phase}).}
\label{fig:MPB}
\end{figure}

Relaxor ferroelectric materials with extremely high 
piezoelectric responses are of great 
interest to both the scientific and industrial
communities~\cite{PZT1}. 
Among those the lead perovskite system 
(1-x)Pb(Zn$_{1/3}$Nb$_{2/3}$)O$_3$-xPbTiO$_3$ (PZN-xPT)  
has been studied extensively, due to its extraordinary piezoelectric 
properties near the morphotropic phase boundary 
(MPB)~\cite{PZN_phase1,PZN_phase2,Universal_phase,PZN_phase,Uesu} (see Inset of 
Fig.~\ref{fig:MPB}). For PZN-xPT, the piezoelectric response reaches maximum 
at $x=8\%$, which is located on the rhombohedral (R) side of the 
MPB~\cite{PZT1}.  As shown in Fig.~\ref{fig:MPB}, the pure PZN system has 
been reported to undergo a 
cubic (C) to rhombohedral (R) phase transition at $T_C=410$~K~\cite{PZN_phase2}.
In the lower portion of Fig.~\ref{fig:MPB}, a schematic of the rhombohedral
angle $\alpha$ vs temperature is shown~\cite{Lebon}.  

The properties of relaxor ferroelectrics have been most commonly described 
in terms of the formation of polar nanoregions (PNR)~\cite{Smolensky,Cross}.
It was found that in relaxor ferroelectric systems, 
PNR start to form at the Burns temperature~\cite{Burns}
and persist below $T_C$. Neutron diffuse scattering 
measurements~\cite{PZN_diffuse,PMN_diffuse} have provided direct evidence
on the existence of PNR. In pure PZN, diffuse scattering intensity due to 
local atomic shifts starts to appear at 450~K and keeps growing 
when the temperature reaches 
below $T_C=410~$K~\cite{PZN_diffuse} (see Fig.~\ref{fig:MPB} for a schematic
of the neutron diffuse scattering intensity vs temperature). 
It was widely accepted that these PNR create random fields under zero 
external fields and 
form ferroelectric micro crystals (MC), which leads to local
rhombohedral symmetry.

The structures of PZN-xPT single crystals have been studied before, but only with 
electric field pre-poled crystals. Measurements with electric field 
along the crystallographic (111) direction~\cite{PZN_Forrester} confirm 
a rhombohedral phase at room temperature in the pre-poled system. Nevertheless, 
no structural study of unpoled PZN single crystals has been reported in the last 
twenty years to our knowledge.
Only until recently, Lebon {\it et al.} studied the single crystal of unpoled PZN using x-ray 
diffraction~\cite{Lebon} with Cu K$_\beta$ x-ray (8.9~keV), 
and observed the rhombohedral splitting for temperature below $T_C$ as 
previously believed. 
 
Recently, Ohwada {\it et al.} performed neutron diffraction measurements  
on the $x=8\%$ PZN-xPT system~\cite{PZN_neutron1} in order to study phase 
transitions in the system with both field-cooling (FC) and zero-field-cooling (ZFC).  
Surprisingly, they did not obtain the rohombohedral (R) phase during the zero-field cooling 
process, but discovered a new, unidentified phase (X). 
This new phase differs from the rhombohedral phase by
having nearly cubic lattice with a slightly different $c-$axis. It is therefore 
interesting to further investigate the ground state of unpoled PZN single crystal
at room temperature to obtain the real crystal structure. 

In our preliminary measurements with 32~keV x-rays on unpole PZN single 
crystals, rhombohedral splittings similar to that obtained in Lebon's paper 
was also observed for peak (111) at room temperature.
However, the bulk property probed by high energy x-ray (67~keV) at
room temperature show a very different structure which is not the widely believed 
rhombohedral phase. Instead, the new ground 
state is an unidentified phase, which 
has a nearly cubic lattice with small tetragonal distortions,
and closely resembles phase X found 
in the $x=8\%$ system. We believe that measurements performed at lower x-ray energies 
only
show the properties of the system near its surface. We will discuss this 
surface phase shortly below, but the main purpose of this paper is to 
describe the new phase X discovered in the bulk of the crystal.

\section{Experiment}

Single crystals of PZN were grown by spontaneous nucleation from high temperature 
solution 
using PbO as flux, as described in Ref.~\onlinecite{PZN_crystal}. 
Two crystal plates of triangular shape (with 
3~mm edges and 1~mm thick) were cut with large surfaces parallel to (111) cubic plane 
from one 
and the same as-grown bulk crystal. The (111) cubic faces of both samples were mirror 
polished 
using a series of diamond pastes down to 3~mm. The (111) cubic faces of one sample were 
sputtered with gold layers. Two gold wires were attached to one corner as electrodes 
using silver paste. 
It was then heated to about 200 °C (above TC), poled by applying an electric field of 
20~kV/cm 
and cooled down to room temperature with the field kept on. The other sample was not 
poled or 
thermally treated, therefore a larger mechanical strain may remain in the crystal.

Previous experiments by Noheda {\it et al.}~\cite{Noheda,Polarization}
show that x-ray diffraction results on PZN-xPT single crystals have a strong
dependence on the surface structures. It is therefore important to use a high
energy x-ray beam preferable in transmission (Laue) mode 
in order to probe the bulk property of the
crystal. Our primary measurements were performed using 67 keV x-rays
at X17B1 beamline of the National 
Synchrotron Light Source (NSLS). The beamline is
equipped with a super-conducting wiggler 
source (4.2 Tesla) providing a
synchrotron x-ray beam with a 
critical energy of 22 keV. 
A monochromatic x-ray beam of 67 keV, with an energy-resolution of
10$^{-4}$ ($\Delta$E/E), was
produced by a sagittal-focusing double-crystal monochromator 
using silicon [311] reflection with both crystals in asymmetric
Laue mode~\cite{zhong01_1}.  The energy of the 
incident beam was calibrated by 4-$\theta$ method using a 
strain-free float-zone silicon crystal.  A perfect-crystal silicon 
analyzer was used with the symmetric Laue reflection [220].  The 
focused beam on the sample was 0.5 mm by 0.5 mm, with a horizontal
and vertical divergence of 50 and 10 micro-radians, respectively.  
The diffraction plane for the sample and analyzer was in the vertical plane. 
For comparison, as discussed later, 
we also performed x-ray diffraction measurements on beam line X-22A at the NSLS 
with lower incident photon energies (32~keV and 10.7~keV).

\section{Skin Effect}

\begin{figure}
\includegraphics[width=\linewidth]{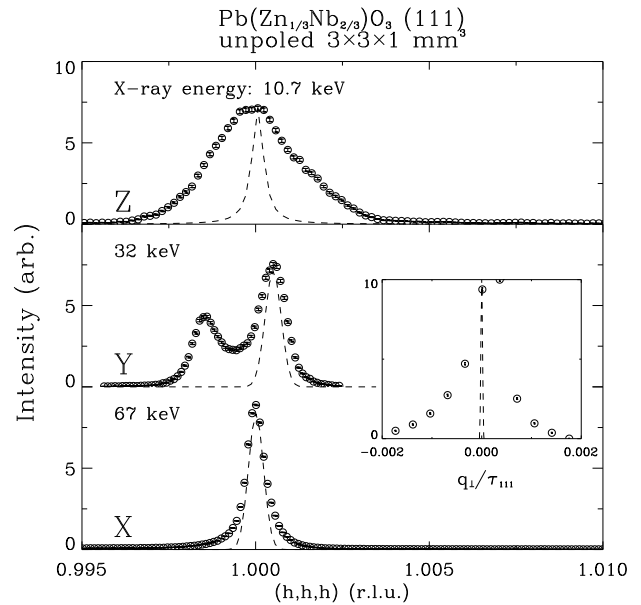}
\caption{Longitudinal scans ($\theta$-$2\theta$ scans) of (111) reflection
on the unpoled PZN single crystal using different x-ray incident energies.
The inset is a transverse scan of (111) taken at x-ray energy of 
67~keV. Units of the horizontal axis are multiples of the pseudocubic 
reciprocal lattice vector (111) $|\tau_{111}|=\sqrt{3}\cdot2\pi/a_0$.}
\label{fig:radial}
\end{figure}

In Fig.~\ref{fig:radial} we show longitudinal scans along 
the pseudocubic (111)
direction of the unpoled PZN single crystal with different x-ray energies.
The instrument resolution curves measured using the (111) reflection of a 
perfect Ge crystal are plotted as dashed lines in the figure.  
The x-ray penetration length and 
lattice parameters extracted from the data 
are listed in Table~\ref{tab:energy}.

For x-ray energy of 10.7~keV, the scan shows a broad peak along 
(111) direction. This can be a result of a collection of micro crystal domains 
contributing to the total diffraction intensity. 
The full width at half maximum (FWHM) of the broad peak is 
$2\Gamma=0.0084$~\AA$^{-1}$, corresponding to an the average size of the micro 
crystal domains of $L=0.94\pi/\Gamma\approx700$~\AA~\cite{Warren}.
When probed with  32~keV x-rays, 
the pseudocubic (111) reflection splits into two peaks.
Both peaks have widths slightly larger than the resolution width. If this 
splitting is related to a rhombohedral distortion, and the two peaks 
are indeed the (111) and ($\bar{1}$11) reflections from different domains,  
the rhombohedral angle $\alpha$ can be determined from the 
positions of the two peaks (Table~\ref{tab:energy}). The result 
($\alpha=89.916^{\circ}$) is in agreement with 
previous measurements (Fig.~\ref{fig:MPB}). 
Yet the scan with 
67~keV shows only one sharp, resolution-limited peak at (111). No rhombohedral
splitting was observed. Also line broadening due to finite (micro) crystal size
effect diminishes at 67~keV.

Lattice parameters of the different phases, marked as X, Y, and Z in 
Fig.~\ref{fig:radial},  can be determined 
by assuming cubic  lattice for phase X and Z, 
and rhombohedral lattice for phase Y (Table~\ref{tab:energy}).
The inconsistency between data sets taken 
at different x-ray energies indicates that the skin effect plays an important
role in our measurements. 
Our diffraction measurements with lower energy 
x-rays were all performed in a reflection mode, because the low energy x-rays 
were not able to penetrate the 1-mm thick sample; 
but with 67~keV x-rays, all the measurements 
were taken in the transmission mode where the x-ray beams were going
through the bulk of the sample. 
At 67~keV, the x-ray penetration length is 
an order of magnitude larger than that at 10.7 and 32~keV, and is 
therefore probing more deeply into the bulk of the crystal. 
Among the different phases
observed, the Y and Z phases are more surface structure related. They have 
distinct features due to inhomogeneities and the microscopic strain caused by boundary 
conditions near the crystal surface. 
The X phase, on the other hand, describes the bulk structure of the 
system at room temperature. 

\begin{table}[h]
\caption{\{111\} reflection of unpoled PZN crystal 
measured at different x-ray energies.}
\begin{ruledtabular}
\begin{tabular}{lccc}
x-ray energy (keV) & 10.7 & 32 & 67 \\
Penetration depth ($\mu$m) & 13 & 59 & 412\\
d-spacing $d_{111}$ (\AA) & 2.3530 &2.3462 &2.3514 \\
d-spacing $d_{\bar{1}11}$ (\AA) & - & 2.3416 & - \\
$a$ (\AA) & 4.0755 & 4.0578 & 4.0728\\
$\alpha$ (deg.) & - & 89.916 & -\\
\end{tabular}
\end{ruledtabular}
\label{tab:energy}
\end{table}

\section{Bulk Properties - Phase X}

\begin{figure}
\includegraphics[width=\linewidth]{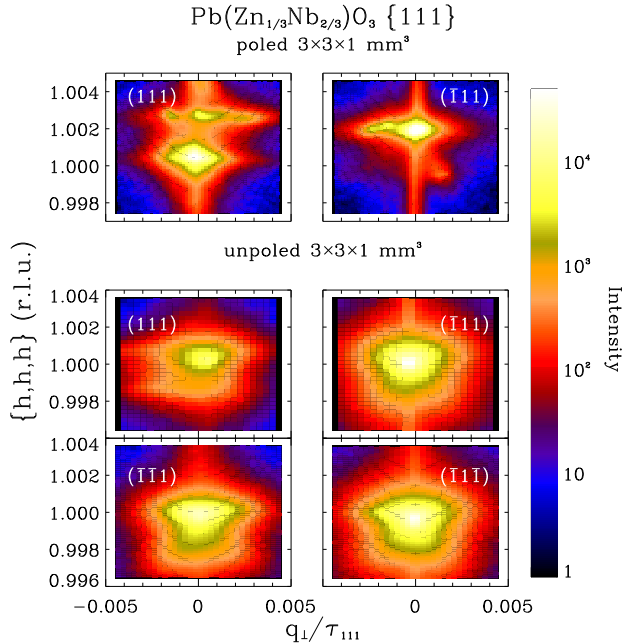}
\caption{Mesh scans taken around pseudocubic \{111\} positions of the 
poled (top frame) and unpoled (bottom frame) PZN single crystals. The x-ray 
energy is 67~keV. The intensity
is plotted in log scale as shown by the scale bar on the right side.  
Units of axes are multiples of the pseudocubic 
reciprocal lattice vector (111) $|\tau_{111}|=\sqrt{3}\cdot2\pi/a_0$. }
\label{fig:mesh111}
\end{figure}

67~keV x-ray diffraction results from the poled PZN single crystal are shown 
in the 
top panel of Fig.~\ref{fig:mesh111}.  Mesh scans in reciprocal space around the 
pseudocubic (111) and ($\bar{1}$11) reflections were performed,
by doing a series of $\theta$ scans at a set of $2\theta$ values 
around the pseudocubic \{111\} reflections. As expected for a rhombohedral
structure, (111) and ($\bar{1}$11) reflections occur at different 2$\theta$s, 
i.e. $d_{111}$ and $d_{\bar{1}11}$ are different due to the rhombohedral 
distortion. Based on the positions of (111) and ($\bar{1}$11) reflections,
the rhombohedral lattice parameters can be obtained: $a=4.0608$~\AA, and 
$\alpha=89.935^\circ$. Weak (111) peak is likely due to a small part of the 
skin which is not fully poled.
The position of the weak peak in the (111) mesh does 
not match exactly with that of the ($\bar{1}$11) peak shown in the other mesh 
scan. 
Based on this splitting in the (111) scan, one can extract a different 
set of rhombohedral lattice parameters: $a=4.0551$~\AA, and 
$\alpha=89.901^\circ$. The likely cause of this discrepancy is the skin effect 
of imperfect poling near the crystal surface.

Results from the unpoled PZN single crystal are, however, entirely 
different. 
In the lower panel of Fig.~\ref{fig:mesh111}, we show four mesh scans at 
the pseudocubic (111),($\bar{1}$11),($\bar{1}\bar{1}$1), and 
($\bar{1}1\bar{1}$) reflections 
for the unpoled PZN single crystal, using high energy (67~keV) x-rays. 
The peak intensities, positions and widths are almost identical in all 
four scans. The lack of diffraction geometry dependence confirms that it is 
the bulk structure, instead of the surface states, that was being probed with 
the 67~keV x-ray beams.
Analysis of the data show that all four \{111\}
reflections have similar d-spacings within $0.05\%$ (Table~\ref{tab:unpoled}).
The four \{111\} mesh scans provide the key evidence that
the structure of the unpoled PZN single crystal is not rhombohedral as 
previously believed. 
To further identify this phase, we performed similar mesh scans at 
the pseudocubic (100), (010), and (001) reflections. Our results 
(Table~\ref{tab:unpoled}) show that the $c$-axis is about 
$0.1\%$ longer than the $a$- and $b$-axes. 
However, we did not observe any peak splitting at the 
\{100\} reflections corresponding to the tetragonal distortion. 
This suggests a tetragonal (T) type distortion but the true symmetry can only
be determined by detailed measurements and analysis of Bragg peak intensities. 
The small tetragonal distortion may be
due to the remaining mechanical stress caused by polishing of the sample.
In this paper we refer to this new unidentified phase of PZN as phase X. 

\begin{table}[h]
\caption{67 keV x-ray diffraction results on unpoled PZN crystal (with 
$0.05\%$ uncertainty).}
\begin{ruledtabular}
\begin{tabular}{cccc}
Reflection & $2\theta$ (deg) & $\delta(2\theta) $ (deg) & $d$ (\AA) \\
\hline
111 & 4.5086 &0.0024&2.3514\\
$\bar{1}$11 & 4.5081 & 0.0022 & 2.3516\\
$\bar{1}\bar{1}$1 & 4.5072 & 0.0035 & 2.3521\\
$\bar{1}$1$\bar{1}$ & 4.5062 & 0.0025 & 2.3526\\
100 & 2.6009 & 0.0020 & 4.0754\\
010 & 2.6011 & 0.0016&4.0750 \\
001 & 2.5975 & 0.0016&4.0801\\
\end{tabular}
\end{ruledtabular}
\label{tab:unpoled}
\end{table}

Our results suggest that the ground state of PZN, namely the phase X, be nearly
cubic with a slight tetragonal distortion of 1.001, characterized by its 
lattice constants shown in Table~\ref{tab:unpoled}. However, the true 
symmetry of this phase is not fully known yet. For this purpose, it is 
essential to study the changes of Bragg intensities through $T_C$ of 410~K. It 
is known that the (200) peak shows a large increase through $T_C$, and it is 
interpreted as the reduction of extinction effect by the formation of micro
ferroelectric state below $T_C$. This is consistent with the large mosaic 
observed (see Inset of Fig.~\ref{fig:radial}) in our measurements 
despite of the sharp longitudinal width. We plan to carry out more detailed 
studies on Bragg line widths, intensities, as well as lattice constants on 
the unpoled PZN crystal at temperatures through $T_C$.

\section{Discussion}

The big puzzle at present is why such a sharp Bragg profile is observed 
below $T_C$ where it has already been demonstrated that a large volume of the 
crystal is occupied by the PNR as indicated by the increasing diffuse 
intensities through $T_C$ (see Inset of Fig.~\ref{fig:MPB}). 
We can visualize a scenario describing our results: 
as shown in Fig~\ref{fig:MPB}, polar nano regions exist in the crystal 
with spontaneous polarizations below $T_C$, but they are incapable of merging 
and forming ferroelectric micro crystals (MC) with rhombohedral symmetry as 
previously expected; and the whole system still retain in average a nearly 
cubic lattice. The average nearly cubic lattice 
gives strong and sharp Bragg peaks at the pseudocubic reflections.
When an electric field is applied to the 
crystal during the cooling process, the PNR align with the external field, thus
making the overall system symmetry rhombohedral instead. 
The important question in this scenario is what is preventing the PNR from 
forming ferroelectric micro crystals.
Recently Hirota {\it et al.}~\cite{PMN_diffuse} demonstrated that a uniform 
displacement of the PNR along their polar direction relative to the 
surrounding unpolarized cubic matrix is required. This phase shifted condense 
mode properly reconciles
the neutron diffuse scattering intensities as well as the soft TO phonon 
intensities in another prototypical lead perovskite 
relaxor system Pb(Mg$_{1/3}$Nb$_{2/3}$)O$_3$ (PMN)~\cite{Ye_Review}.
We believe that this uniform phase shift may be playing a key role in 
answering the 
previously raised question. When the PNR are shifted with respect to their
surrounding environments, the lattice distortion 
could cause a higher energy barrier for the PNR
to reorient, therefore ``freezing'' the PNR at their local polarizations; and 
makes it harder for them to merge into 
larger polar crystals which eventually lead to a long range 
ferroelectric phase below $T_C$. 

It is also important to realize that in the other relaxor system PMN, similar
behavior has been observed below $T_C$. It was also found that with zero electric 
field cooling,
PMN undergoes a transition into a random-bond-driven glassy state instead of a 
ferroelectric state~\cite{PMN_Bobnar}.
This system has been intensively 
investigated in recent years~\cite{PMN_neutron,PMN_neutron2}, and a more 
detailed description on this compound can be found in the review article
by Ye~\cite{Ye_Review}. Until
recently the accepted picture was that PMN remain cubic through its apparent
$T_C$ of 220~K. The recent neutron scattering paper by 
Wakimoto {\it et al.}~\cite{Waki1} has changed the picture. The soft mode
energy square increases linearly with temperature below $T_C$, as expected 
for ordered ferroelectric phase. Our discovery of phase X in PZN hints that 
the ``cubic'' phase of PMN below $T_C$ might be the phase X. Further 
experiments are needed to confirm this speculation as well as the phase X in 
the rhombohedral region of PMN-xPT~\cite{PMN_phase}.

\section{Conclusions}

In summary, we have performed detailed high energy x-ray diffraction 
measurements on single crystals of the relaxor ferroelectric PZN. 
A new phase (X) was discovered at room temperature having nearly cubic
lattice with a slight tetragonal distortion.
At  temperature below $T_C$, the system does 
not transform into 
a long range ferroelectric phase. Instead, the system is best described by a 
collection of local polar nanoregions embedded in an overall 
nearly cubic lattice. The PNR  do not merge into 
large polar domains with long range ferroelectric order. 
The discovery of phase X in both pure PZN and $8\%$ PZN-PT~\cite{PZN_neutron1}
indicates that the 
ground state of the previously believed rhombohedral (R) side of the MPB 
need to be modified. We also believe that a similar phase exists in the other 
prototypical lead perovskite relaxor ferroelectric PMN as well. 
This disordered ground state would change the whole 
picture of the phase diagram of the relaxor system, 
and provide new exciting challenges for future theoretical and experimental 
work. In addition, our preliminary studies on the skin effect
of the crystal show very interesting results. Different profiles (Y and Z) have 
been observed from the surface of the crystal under different conditions. 
These surface states are very important but not fully understood yet. 
Further studies are being carried out to better identify 
these surface states and their origins. 

\section{Acknowledgments}

We would like to thank W.~Chen, D.~Cox, J.~Hill, K.~Hirota, B.~Noheda, 
B.~Ocko, K.~Ohwada, and R.~Werner for stimulating discussions. 
Financial support from the U.S. Department of Energy under contract 
No.~DE-AC02-98CH10886, U.S. Office of Naval Research Grant No.~N00014-99-1-0738,
and the Natural Science and Research Council of Canada 
(NSERC) is also gratefully acknowledged.

\end{document}